# From old wars to new wars and global terrorism

N. Johnson[1,7], M. Spagat[2,7], J. Restrepo[3,7], J. Bohórquez[4], N. Suárez[5,7], E. Restrepo[6,7], and R. Zarama[4]

*[1] Department of Physics, University of Oxford, Oxford, U.K.*

*[2] Department of Economics, Royal Holloway College, University of London, Egham, U.K.*

*[3] Department of Economics, Universidad Javeriana, Bogotá, Colombia*

*[4] Department of Industrial Engineering, Universidad de los Andes, Bogotá, Colombia*

*[5] Department of Economics, Universidad Nacional de Colombia, Bogotá, Colombia*

*[6] Department of Economics, Universidad de Los Andes, Bogotá, Colombia*

*[7] CERAC, Conflict Analysis Resource Center, Bogotá, Colombia*

## ABSTRACT

**Even before 9/11 there were claims that the nature of war had changed fundamentally[1]. The 9/11 attacks created an urgent need to understand contemporary wars and their relationship to older conventional and terrorist wars, both of which exhibit remarkable regularities[2-6]. The frequency-intensity distribution of fatalities in "old wars", 1816-1980, is a power-law with exponent 1.80(9)[2].[i] Global terrorist attacks, 1968-present, also follow a power-law with exponent 1.71(3) for G7 countries and 2.5(1) for non-G7 countries[5]. Here we analyze two ongoing, high-profile wars on opposite sides of the globe - Colombia and Iraq. Our analysis uses our own unique dataset for killings and injuries in Colombia, plus publicly available data for civilians killed in Iraq. We show strong evidence for power-law behavior *within* each war. Despite substantial differences in contexts and data coverage, the power-law coefficients for *both* wars are tending toward 2.5, which is a value characteristic of non-G7 terrorism as opposed to old wars. We propose a plausible yet analytically-solvable model of modern insurgent warfare, which can explain these observations.**

**Correspondence:** M.Spagat@rhul.ac.uk (M. Spagat)    n.johnson@physics.ox.ac.uk (N. Johnson)

---

[i] Numbers in parentheses give the standard error on the trailing figure in each case.

In two celebrated papers[3,4] Lewis Richardson showed that war casualties follow a power law distribution, i.e. the probability that a given war has $x$ victims, $p(x)$, is equal to $Cx^{-\alpha}$ over a reasonably wide range of $x$, with $C$ and $\alpha$ positive coefficients. This in turn implies that a graph of $\log[P(X \geq x)]$ vs. $\log(x)$ will be a straight line over this range of $x$, with negative slope $\alpha - 1$.[ii] These results were updated recently[6] to show that interstate wars, 1820-1997, obey a power law. Each data point is a casualty count for an entire war in these studies. Casualty numbers in global terrorist events, 1968 to present, also obey power laws where in this case each data point is a terrorist attack[5].

While many people believe that 9/11 fundamentally changed the nature of warfare, some analysts had discerned new wars emerging even before this disaster[1]. Thomas Hammes views "fourth generation wars", as trenchantly exposited by Mao Tse-tung, as the prevalent form of contemporary warfare[7,8]. These are conflicts in which incumbents with overwhelming military and economic superiority face extremely patient insurgents seeking to break their enemies' political will through persistent and demoralizing attacks. The phenomenon covers numerous well-known cases including Viet Nam, Iraq, Afghanistan, Israel-Palestine and Al-Qaeda[7]. Thus, fourth generation warfare encompasses both global terrorism[5] plus a variety of civil and/or international wars as commonly understood.

Here our contribution is threefold. First, we analyze detailed daily data for two specific ongoing wars in Colombia and Iraq and find that both obey power laws. Thus, we extend Richardson's fundamental insight into the micro world of single conflicts. Second, we show that the power-law coefficients for both wars are drifting strikingly close to the global terrorism coefficient for non-G7 countries[5]. Thus, at least these two examples of

---

[ii] We will refer to $P(X \geq x)$ as the cumulative distribution obtained from $p(x)$.

modern warfare increasingly resemble both each other and global terrorism in non-G7 countries. This finding resonates strongly with the notion of the rise of fourth generation warfare[7]. Third we propose a micro conflict model that can explain our results.

Figure 1 shows log-log plots of the fraction of all recorded events for that particular war with $x$ or more victims, $P(X \geq x)$, versus $x$. For Colombia we are able to work with the very broad measure of all conflict-related killings plus injuries. For the Iraq data we work with killings of civilians as provided by the Iraq Body Count Project. The straight lines over long ranges in Figure 1 suggest that both these wars follow power laws. The Colombia data displays an extraordinary fit for a social science application while the Iraq data also fits well except for a bulge in the 150 to 350 range. Since we have many more Colombia events than Iraq ones, the superiority of the Colombia curve is not surprising. Nevertheless, the rest of the Iraq curve fits well enough so as to suggest that we should expect more events in the 150-200 range in the future. (The inset to Figure 1 shows a shortage of events in the 150-200 range. The cumulative distribution therefore exhibits a bulge, which eventually disappears around 350).

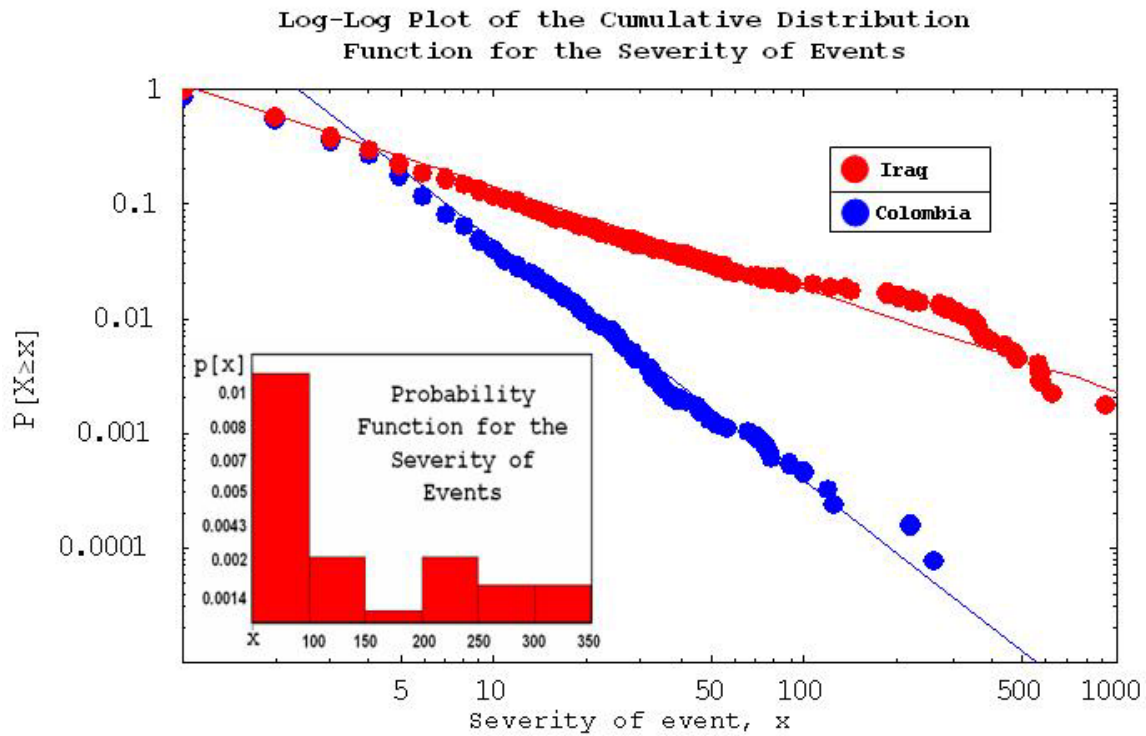

Figure 1

**Figure 1  Log-log plots of cumulative distributions** $P(X \geq x)$ **describing the total number of events with severity greater than** $x$ **, for the ongoing wars in Iraq (red) and Colombia (blue).  For Iraq, the severity is taken to be the lower estimate of civilian deaths from www.iraqbodycount.com.  For Colombia, the severity is taken to be the total number of deaths plus injuries from the CERAC dataset[9].  Each line indicates the most likely power law that fits the data (see text).  The inset shows a histogram of the Iraq data set and points to a shortage of attacks with severity in the range 150-200; this shortage creates the bulge in the Iraq line in the main figure.**

Using well-established methods[2], as explained in the Methods and Supplementary Information sections, we have verified that each cumulative distribution in Figure 1 satisfies a power-law relationship over a very wide range. We also find robust power-law behavior for data collected over smaller time-windows, as discussed below, and have hence deduced the evolution of the power-law coefficient $\alpha$ over time by sliding this time-window through the data-series. Figure 2 shows these empirically-determined $\alpha$ values as a function of time for both conflicts. The $\alpha$ values in both cases are tending toward 2.5, which is the coefficient for global terrorism in non-G7 countries. The implication is that *both* these wars *and* global non-G7 terrorism are beginning to share a similar underlying structure. This finding is consistent with the idea of the increasing prevalence of fourth generation warfare[7]. The Methods and Supplementary Information sections provide details of the tests we performed to verify the robustness of our results.

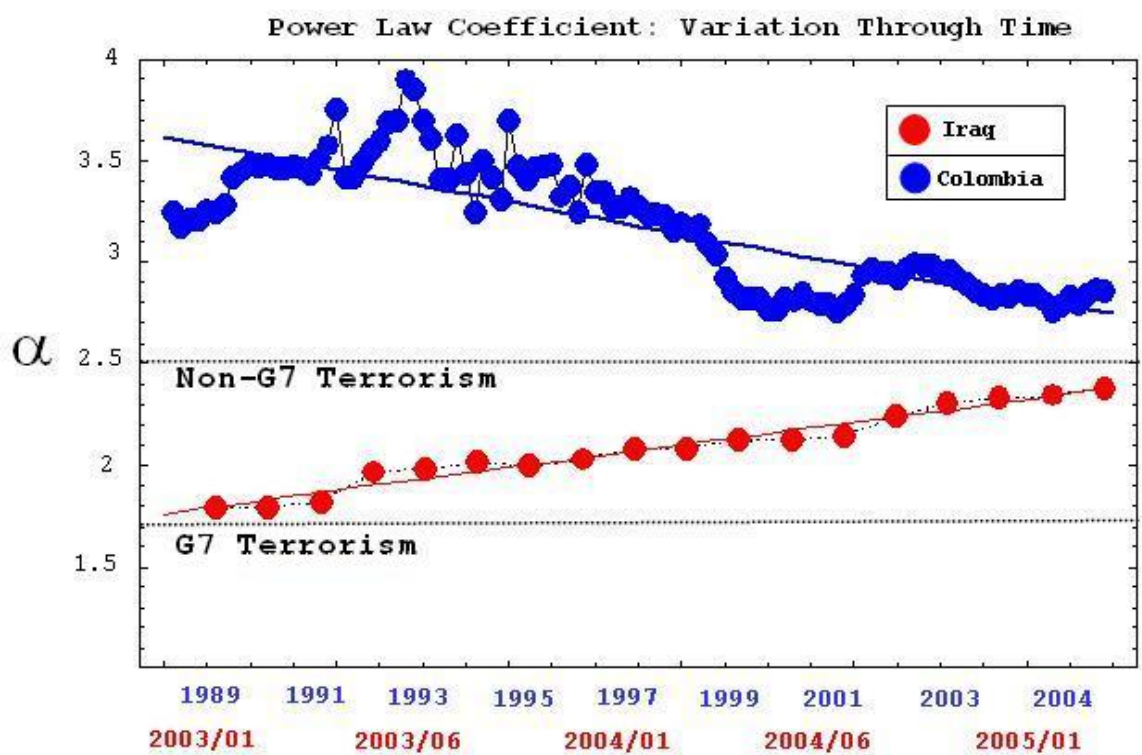

Figure 2

**Figure 2  The variation through time of the power law coefficient $\alpha$ for Iraq (red) and Colombia (blue).  The straight lines are fits through these points, and suggest a common value of approximately 2.5 for both wars in the near future. The values for G7 and non-G7 terrorism are also shown[5]. See text for details of how the variation through time of $\alpha$ is calculated.**

There is a need for a model which can explain this common value of $\alpha \approx 2.5$. Standard physical mechanisms for generating power laws make little sense in the context of Colombia or Iraq[2]. One might instead guess that casualties would arise in rough proportion to the population sizes of the places where insurgent groups attack: given that city populations may follow a power law[2], it is conceivable that this would also produce power laws for the severity of attacks. However, we have tested this hypothesis against our Colombia data and it is resoundingly rejected.

Instead, we have developed a new model of modern insurgent warfare. As shown in Figure 3, and explained in detail in the Supplementary Information section, our model assumes that the insurgent force operates as a collection of fairly self-contained units, which we call 'attack units'. Each attack unit has a particular 'attack strength' characterizing the average number of casualties arising in an event involving this attack unit. As time evolves, these attack units either join forces with other attack units (i.e. coalescence) or break up (i.e. fragmentation). Eventually this on-going process of coalescence and fragmentation reaches a dynamical steady-state which is solvable analytically, yielding $\alpha = 2.5$. This value is in remarkable agreement with the $\alpha$ values to which both Colombia and Iraq appear to be tending (recall Figure 2). It also suggests that similar distributions of attack units might be emerging in both Colombia and Iraq, with each attack unit in an ongoing state of coalescence and fragmentation. Our model also offers the following interpretation for the dynamical evolution of $\alpha$ observed in Figure 2. The Iraq war began as a conventional confrontation between large armies, but continuous pressure applied to the Iraqis by coalition forces has fragmented the insurgency into a structure in which smaller attack units, characteristic of non-G7 global terrorism, now predominate. In Colombia, on the other hand, the guerrillas in the early 1990's had even less ability than global terrorists to coalesce into high-impact units but have gradually been acquiring comparable capabilities.

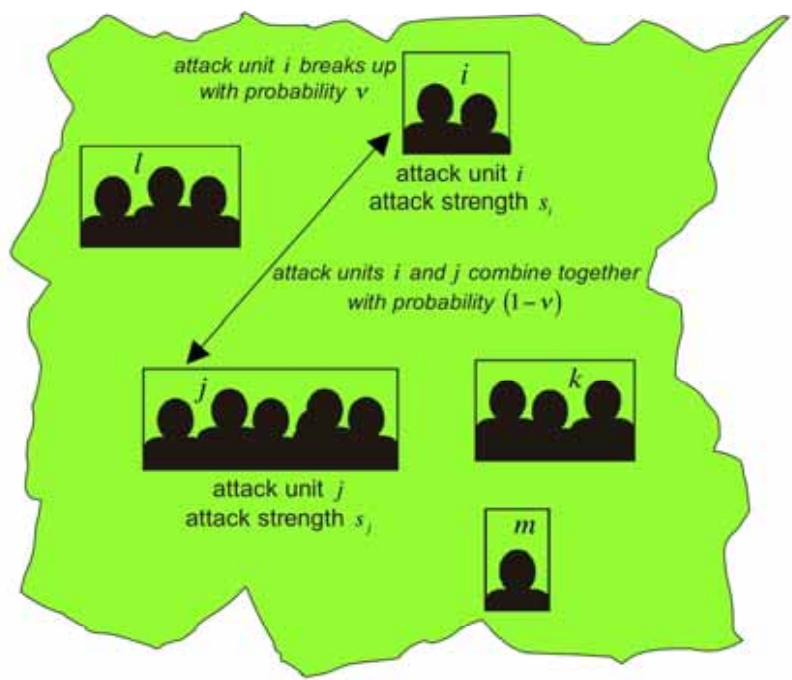

Figure 3

**Figure 3  Our analytically-solvable model describing modern insurgent warfare. The insurgent force comprises *attack units*, each of which has a particular *attack strength*. The total attack strength of the insurgent force is being continually re-distributed through a process of coalescence and fragmentation.  Mathematical details are provided in the Supplementary Information section.**

More generally, our results – combined with those of Clauset and Young, and Richardson – suggest that there are power laws between wars, power laws for global terrorism and power laws within contemporary wars. That is, power laws have an extraordinary range of applicability to human conflict, both on the large and the small scale. In addition, our finding that the statistical patterns of the intra-war events in Colombia and Iraq appear to be trending toward the same value as global terrorist incidents in non-G7 countries (i.e. $\alpha = 2.5$) suggests that such global terrorist incidents can themselves be viewed as intra-war events within some larger, on-going yet ill-defined "global war". This leaves open the possibility that the spatio-temporal correlations between events within a particular war, are related to those at play in global terrorism. We leave this intriguing discussion to a later publication.

## Methods

### Data Sources

We make extensive use of our own CERAC dataset for Colombia[9] plus publicly available data on Iraq (www.iraqbodycount.org). The CERAC data builds on primary source compilations of violent events by Colombian human rights NGO's and from local and national press reports. We distil from this foundation all the clear conflict events, i.e., those that have a military effect and reflect the actions of a group participating in the armed conflict. For each event we record the participating groups, the type of event (massacre, bombing, clash, etc.), the location, the methods used and the number of killings and injuries of people in various categories (guerrillas, civilians, etc.). This data set covers the years 1988-2004 and includes 20,251 events. The Iraq Body Count Project monitors the reporting of more than 30 respected online news sources, recording only events reported by at least two of them. For each event they log the date, time, location, target, weapon, estimates of the minimum and maximum number of civilian deaths and the sources of the information.

The concept of civilian is broad, including, for example, policemen. The list of events, posted online, covers the full range of war activity, including suicide bombings, roadside bombings, US air strikes, car bombs, artillery strikes and individual assassinations. The data set covers the period from 2003 to the present and includes 1,746 events.

**Power Law Calculation**

First we used a Kolmogorov-Smirnov goodness-of-fit test to select $x_{min}$, the smallest value for which the power law is thought to hold. The formula $\alpha = 1 + n \left[ \sum_{i=1}^{n} \ln\left(x_i / x_{min}\right) \right]^{-1}$ then estimates the power-law exponent while jackknife resampling estimates the error in $\alpha$. To check these results, we then estimated $\alpha$ using least-square regression on the observations above $x_{min}$. For both the Iraq and the Colombia data we obtained nearly identical point estimates of very high significance, with nearly null $p$-values using White-heteroskedasticity-corrected robust standard errors. We then performed robustness checks by excluding outliers and high-leverage observations from the regressions, finding only marginal changes in parameter estimates.

**Variation through time of $\alpha$**

We apply the above procedures by varying $x_{min}$ for each estimate and also by using a fixed $x_{min}$ for all the estimates, and find no significant differences. The Colombian coefficients are calculated for two-year intervals displaced every 50 days. The Iraq coefficients are calculated for one-year intervals displaced every 30 days. The differences in calculation procedures were necessitated by the relatively shorter run of the Iraq data compared to the Colombia data.

**Acknowledgements** The Department of Economics of Royal Holloway College provided funds to build the CERAC database. J. Restrepo acknowledges funding from Banco de la República, Colombia. J.C. Bohórquez acknowledges funding from the Department of Industrial Engineering of Universidad de los Andes.

# Supplementary Information

## PART 1:   Detailed discussion of the model introduced in the paper

Here we provide details of the model of modern insurgent warfare, which we introduced in the main paper. Our goal is to provide a plausible model to explain (i) why power-law behaviour is observed in the Colombia and Iraq wars, and (ii) why the power-law coefficients for the Colombia and Iraq wars should both be heading toward a value of 2.5. In other words, *why should a modern war such as that currently underway in Colombia or Iraq, produce power-law behaviour and why should the value of 2.5 emerge as a power-law coefficient?*

Our model bears some similarity to a model of herding by Cont and Bouchaud[iii], and is a direct adaptation of the Eguiluz-Zimmerman model of herding in financial markets[iv]. The analytical derivation which we present, is an adaptation of earlier formalism laid out by D'Hulst and Rodgers[v], and also draws heavily on the material in the book *Financial Market Complexity* by Neil F. Johnson, Paul Jefferies and Pak Ming Hui (Oxford University Press, 2003). One of us (NFJ) is extremely grateful to Pak Ming Hui for detailed correspondence about the Eguiluz-Zimmerman model of financial markets, the associated formalism, and its extensions – and also for discussions involving the present model.

As suggested by Figure 3 in the paper, our model is based on the plausible notion that the total attack capability of an insurgent force in 'fourth-generation' warfare, is being

[iii] R. Cont and J.P. Bouchaud, Macroeconomic Dynamics **4**, 170 (2000)

[iv] V.M. Eguiluz and M.G. Zimmerman, Phys. Rev. Lett. **85**, 5659 (2000)

[v] R. D'Hulst and G.J. Rodgers, Eur. Phys. J. B **20**, 619 (2001). See also Y. Xie, B.H. Wang, H. Quan, W. Yang and P.M. Hui, Phys. Rev. E **65**, 046130 (2002).

continually re-distributed. Based on our intuition about such guerilla-like wars, we consider the insurgent force to be made up of ***attack units*** or ***cells*** which have certain ***attack strength*** (see below for a detailed discussion). One might expect that the total attack strength for the entire insurgent force would change slowly over time.  At any particular instant, this total attack strength is distributed (i.e. partitioned) among the various attack units -- moreover the composition of these attack units, and hence their relative attack strengths, will evolve in time as a result of an on-going process of coalescence (i.e. combination of attack units) and fragmentation (i.e. breaking up of attack units). Such a process of coalescence and fragmentation is realistic for an insurgent force in a guerilla-like war, and will be driven by a combination of planned decisions and opportunistic actions by both the insurgent force and the incumbent force. For example, separate attack units might coalesce prior to an attack, or an individual attack unit might fragment in response to a crackdown by the incumbent force. Here we will model this process of coalescence and fragmentation as a stochastic process.

Each attack unit carries a specific label $i, j, k, \ldots$ and has an attack strength denoted by $s_i, s_j, s_k, \ldots$ respectively. We start by discussing what we mean by these definitions:

***attack unit*** or ***cell***: Here we have in mind a group of people, weapons, explosives, machines, or even information, which organizes itself to act as a single unit. In the case of people, this means that they are probably connected by location (e.g. they are physically together) or connected by some form of communications systems. In the case of a piece of equipment, this means that it is readily available for use by members of a particular group. The simplest scenario is to just consider people, and in particular a group of insurgents which are in such frequent contact that they are able to act as a single group. However we emphasize that an attack unit may also consist of a combination of people and objects – for example, explosives plus a few people, such as the case of suicide bombers. Such an attack unit, while only containing a few people, could have a high attack strength. In addition, information could also be a valuable part of an attack unit. For example, a lone suicide bomber who knows when a certain place will be densely populated (e.g. a military canteen at lunchtimes) and who knows how to get into such a place unnoticed, will also represent an attack unit with a high attack strength.

***attack strength***: We define the attack strength $s_i$ of a given attack unit $i$, as the average number of people who are typically injured or killed as the result of an event involving attack unit $i$. In other words, a typical event (e.g. attack or clash) involving group $i$ will lead to the injury or death of $s_i$ people. This definition covers both the case of one-sided attacks by attack unit $i$ (since in this case, all casualties are due to the presence of attack unit $i$) and it also covers two-sided clashes (since presumably there would have been no clash, and hence no casualties, if unit $i$ had not been present).

We take the sum of the attack strengths over all the attack units (i.e. the total attack strength of the insurgent force) to be equal to $N$. From the definition of attack strength, it follows that $N$ represents the maximum number of people which would be injured or killed in an event, on average, if the entire insurgent force were to act together as a single attack unit. Mathematically, $\sum_{i,j,k,...} s_i = N$. For any significant insurgent force, one would expect $N >> 1$. The power-law results that we will derive do not depend on any particular choice of $N$. In particular, the power-law result which is derived in this Supplementary Information section concerning the average number $n_s$ of attack units having a given attack strength $s$, is invariant under a global magnification of scale (as are all power-laws).

The model therefore becomes, in mathematical terms, one in which this total attack strength $N$ is dynamically distributed among attack groups as a result of an ongoing process of coalescence and fragmentation. As a further clarification of our terminology, we will now discuss the two limiting cases which we classify as the 'coalescence' and 'fragmentation' limits for convenience:

- 'Coalescence' limit: Suppose the conflict is such that all the attack units join together or *coalesce* into a single large attack unit. This is the limit of complete coalescence and would correspond to amassing all the available combatants and weaponry in a single place – very much like the armies of the past would amass their entire force on the field of battle. Hence there is one large attack unit, which we label as $i$ and which has an attack strength $N$. All other attack units disappear. Hence $s_i \rightarrow N$. This 'coalescence' limit has the *minimum* possible number of attack units (i.e. one) but the *maximum* possible attack strength (i.e. $N$) in that attack unit.

- 'Fragmentation' limit: Suppose the conflict is such that all the attack units *fragment* into ever smaller attack units. Eventually we will have all attack units having attack strength equal to one. Hence $s_i \rightarrow 1$ for all $i = 1, 2, \ldots, N$. This would correspond to all combatants operating essentially individually. This 'fragmentation' limit has the *maximum* possible number of attack units (i.e. $N$) but the *minimum* possible attack strength per attack unit (i.e. one).

In practice, of course, one would expect the situation to lie between these two limits. Indeed, it seems reasonable to expect that these attack units and their respective attack strengths, will evolve in time within a given war. Indeed, one can envisage that these attack units will occasionally either break up into smaller groups (i.e. smaller attack units) or join together to form larger ones. The reasons are plentiful why this should occur: for example, the opposing forces (e.g. the Colombian Army in Colombia, or Coalition Forces in Iraq) may be applying pressure in terms of searching for hidden insurgent groups. Hence these insurgent groups (i.e. attack units) might either decide, or be forced, to break up in order to move more quickly, or in order to lose themselves in the towns or countryside.

Hence attack units with different attack strengths will continually mutate via coalescence and fragmentation yielding a 'soup' of attack units with a range of attack strengths. At any one moment in time, this 'soup' corresponds mathematically to partitioning the total $N$ units of attack strength which the insurgent army possesses. The analysis which we now present suggests that the current states of the guerilla/insurgency wars in Colombia and Iraq both correspond to the steady-state limit of such an on-going coalescence-fragmentation process. It also suggests that such a process might also underpin the acts of terrorism in non-G7 countries, and that such terrorism is characteristic of some longer-term 'global war'.

Against the backdrop of on-going fragmentation and coalescence of attack units, we suppose that each attack unit has a given probability $p$ of being involved in an event in a given time-interval, regardless of its attack strength. For example, $p$ could represent the probability that an arbitrarily chosen attack unit comes across an undefended target – or vice versa, the probability that an arbitrarily chosen attack unit finds itself under attack. In these instances, $p$ should be relatively insensitive to the actual attack strength of the attack unit involved: hence the results which we shall derive for the distribution of attack strengths, should also be applicable to the distribution of events having a given severity. When obtaining our analytic and numerical results, we assume that the war has been underway for a long time and hence some kind of steady-state has been reached. This latter assumption is again plausible for the wars in Colombia and Iraq.

Given the above considerations, it follows that if there are, on average, $n_s$ attack units of a given attack strength $s$, then the average number of events involving an attack unit of attack strength $s$ will be proportional to $n_s$. We assume, quite realistically, that only one insurgent attack group participates in a given event. For example, an attack in which 10 people were killed is necessarily due to an attack by a unit of attack strength 10. In particular, it could not be due to two separate but simultaneous attacks by a unit of strength 6 and a unit of strength 4 (i.e. 6+4=10). Hence the number of events in which $s$ people were killed and/or injured, is just proportional to $n_s$. In other words, the histogram, and hence power-law, that we will derive for the dependence of $n_s$ on $s$, will also describe the number of events with $s$ casualties versus $s$. Indeed, if we consider that an event will typically have a duration of $T$, and that there will only be a few such events in a given interval $T$, then these results should also appear similar to the distribution describing the number of intervals of duration $T$ in which there were $s$ casualties, versus $s$. This is indeed what we have found in our analysis of the empirical data.

Given these considerations, our task of analyzing and deducing the average number of events with $s$ casualties versus $s$ over a given period of time, becomes equivalent to the task of analyzing and deducing the average number $n_s$ of attack units of a given attack strength $s$ in that same period of time. This is what we will now calculate. We will start by considering a mechanism for coalescence and fragmentation of attack groups, before then finally deducing analytically the corresponding power-law behaviour and hence deducing a power-law coefficient equal to 2.5.

Consider an arbitrary attack unit $i$ with attack strength $s_i$. At any one instant in time, labelled $t$, we assume that this attack unit may either:

a) fragment (i.e. break up) into $s_i$ attack units of attack strength equal to 1. This feature aims to mimic an insurgent group which decides, either voluntarily or involuntarily, to split itself up (e.g. in order to reduce the chance of being captured and/or to mislead the enemy).

b) coalesce (i.e. combine) with another attack unit $j$ of attack strength $s_j$, hence forming a single attack unit of attack strength $s_i + s_j$. This feature mimics two insurgent groups finding each other by chance (e.g. in the Colombian jungle) or deciding via radio communication to meet up and join forces.

To implement this fragmentation/coalescence process at a given timestep, we choose an attack unit $i$ at random but with a probability which is proportional to its attack strength $s_i$. With a probability $\nu$, this attack unit $i$ with attack strength $s_i$ *fragments* into $s_i$ attack units with attack strength 1. A justification for choosing attack unit $i$ with a probability which is proportional to its attack strength, is as follows: attack units with higher attack strength are likely to be bigger and hence will either run across the enemy more and/or be more actively sought by the enemy. By contrast, with a probability $(1 - \nu)$, the chosen attack unit $i$ instead *coalesces* with another attack unit $j$ which is chosen at random, but again with a probability which is proportional to its attack strength $s_j$. The two attack units of attack strengths $s_i$ and $s_j$ then combine to form a bigger attack unit of attack strength $s_i + s_j$. The justification for choosing attack unit $j$ for coalescence with a probability which is proportional to its attack strength, is as follows: it is presumably risky to combine attack units, since it must involve at least one message passing between the two units in order to coordinate their actions. Hence it becomes increasingly less worthwhile to combine attack units as the attack units get smaller.

This model is thus characterized by a single parameter $\nu$. The set up of the model is shown schematically in the figure at the front of this Supplementary Information section, and in Figure 3 of the paper. The connectivity among the attack units is driven by the dynamics of the model. For very small $\nu$ (i.e. much less than 1), the attack units steadily coalesce. This leads to the formation of large attack units. In the other limit of $\nu \rightarrow 1$, the system consists of many attack units with attack strength close to 1. A value of $\nu = 0.01$ corresponds to about one fragmentation in every 100 iterations. In what follows, we assume that $\nu$ is small since the process of fragmentation should not be very frequent for any insurgent force which is managing to sustain an ongoing war. Indeed if such fragmentation were very frequent, then this would imply that the insurgents were being so pressured by the incumbent force that they had to fragment at nearly every timestep. Hence that particular war would not last very long. It turns out that infrequent fragmentations are sufficient to yield a steady-state process, and will also yield the power-law behaviour which we observe for Colombia and Iraq.

A typical result obtained from numerical simulations, for the distribution of $n_s$ versus attack strength $s$ in the long-time limit (i.e. steady-state), is shown below in terms of $n_s/n_1$ :

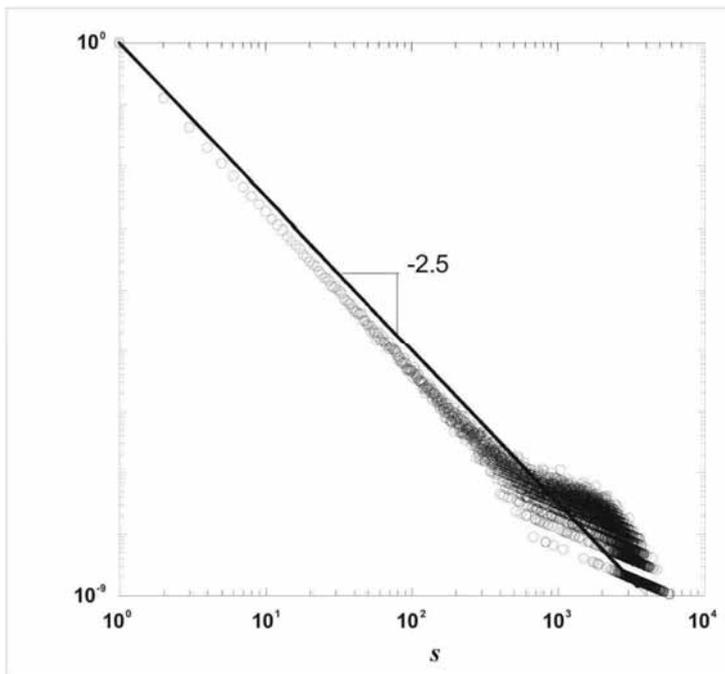

Supplementary Figure 1: Log-log plot of the number of attack units with attack strength $s$, versus attack strength $s$. Here $N = 10,000$ and $v = 0.01$. The results are obtained from a numerical simulation of the model. The initial conditions of this numerical simulation are such that all attack units have size 1. As time evolves, these attack units undergo coalescence and fragmentation as described in the text. In the long-time limit, the system reaches a steady state with a power-law dependence as shown in the figure, and with an associated power-law coefficient of 2.5 (i.e. 5/2). The deviation from power-law behaviour at large $s$ is simply due to the finite value of $N$: since there can be no attack unit with an attack strength greater than $N$, the finite size of $N$ distorts the power-law as $s$ approaches $N$.

We now provide an *analytic* derivation of the observed power-law behaviour, and specifically the power-law coefficient 2.5, in the steady-state (i.e. long-time) limit.

One could write a dynamical equation for the evolution of the model with different levels of approximation. For example, one could start with a microscopic description of the system by noting that at any moment in time, the entire insurgent army can be described by a partition $\{l_1, l_2, \ldots, l_N\}$ of the total attack strength $N$ into $N$ attack units. Here $l_s$ is the number of attack units of attack strength $s$. For example $\{0, 0, \ldots, 1\}$ corresponds to the extreme coalescence case in which all the attack strength is concentrated in one big attack unit. By contrast, $\{N, 0, \ldots, 0\}$ corresponds to the case of extreme fragmentation in which all the attack units have attack strength of 1 (i.e. there are $N$ attack units of attack strength 1). Clearly, the total amount of attack strength is conserved $\sum_{i=1}^{N} i l_i = N$. All that happens is that the way in which this total attack strength $N$ is *partitioned* will change in time.

In principle, the dynamics could be described by the time-evolution of the probability function $p[l_1, l_2, \ldots, l_N]$: in particular, taking the continuous-time limit would yield an equation for $dp[l_1, l_2, \ldots, l_N]/dt$ in terms of transitions between partitions. For example, the fragmentation of an attack unit of attack strength $s$ leads to a transition from the partition $\{l_1, \ldots, l_s, \ldots, l_N\}$ to the partition $\{l_1 + s, \ldots, l_s - 1, \ldots, l_N\}$. For our purposes, however, it is more convenient to work with the *average* number $n_s$ of attack units of attack strength $s$, which can be written as $n_s = \sum_{\{l_1, \ldots, l_N\}} p[l_1, \ldots, l_s, \ldots, l_N] \cdot l_s$. The sum is over all possible partitions. Since $p[l_1, \ldots, l_N]$ evolves in time, so does $n_s[t]$. After the transients have died away, the system is expected to reach a steady-state in which $p[l_1, \ldots, l_N]$ and $n_s[t]$ become time-independent. The time-evolution of $n_s[t]$ can be written down either by intuition, or by invoking a mean-field approximation to the equation for $dp[l_1, l_2, \ldots, l_N]/dt$. Taking the intuitive route, one can immediately write down the following dynamical equations in the continuous-time limit:

$$\frac{\partial n_s}{\partial t} = -\frac{\nu s n_s}{N} + \frac{(1-\nu)}{N^2} \sum_{s'=1}^{s-1} s' n_{s'} (s-s') n_{s-s'} - \frac{2(1-\nu) s n_s}{N^2} \sum_{s'=1}^{\infty} s' n_{s'} \qquad \text{for } s \geq 2 \qquad (0.1)$$

$$\frac{\partial n_1}{\partial t} = \frac{\nu}{N} \sum_{s'=2}^{\infty} (s')^2 n_{s'} - \frac{2(1-\nu)n_1}{N^2} \sum_{s=1}^{\infty} s' n_s \qquad (0.2)$$

The terms on the right-hand side of Equation (0.1) represent all the ways in which $n_s$ can change. The first term represents a decrease in $n_s$ due to the fragmentation of an attack unit of attack strength $s$: this happens only if an attack unit of attack strength $s$ is chosen and if fragmentation then follows. The former occurs with probability $s\,n_s/N$ (see earlier discussion) and the latter with probability $\nu$. The second term represents an increase in $n_s$ as a result of the merging of an attack unit of attack strength $s'$ with an attack unit of attack strength $(s - s')$. The third term describes the decrease in $n_s$ due to the merging of an attack unit of attack strength $s$ with any other attack unit. For the $s = 1$ case described by Equation (0.2), the chosen attack unit remains isolated; thus Equation (0.2) does not have a contribution like the first term of Equation (0.1). The first term which appears in Equation (0.2) reflects the increase in the number of attack units of attack strength equal to 1, due to fragmentation of an attack unit. Similarly to Equation (0.1), the last term of Equation (0.2) describes the merging of an attack unit with attack strength 1, with an attack unit of any other attack strength. Equations (0.1) and (0.2) are so-called 'master equations' describing the dynamics within the model. Note that for simplicity, we are only considering fragmentation into attack units of attack strength 1. However this could be generalized – indeed, we will look at more general fragmentations in future publications.

In the long-time steady state limit, Equations (0.1) and (0.2) yield:

$$s\,n_s = \frac{(1-\nu)}{(2-\nu)N} \sum_{s'=1}^{s-1} s' n_{s'} (s - s') n_{s-s'} \quad \text{for } s \geq 2 \qquad (0.3)$$

$$n_1 = \frac{\nu}{(2-\nu)N} \sum_{s'=2}^{\infty} (s')^2 n_{s'} \qquad (0.4)$$

Equations of this type are most conveniently treated using the general technique of 'generating functions'. As the name suggests, these are functions which can be used to generate a range of useful quantities. Consider

$$G[y] = \sum_{s=0}^{\infty} s' n_s \, y^s \qquad (0.5)$$

where $y = e^{-\omega}$ is a parameter. Note that $s \, n_s / N$ is the probability of finding an attack unit of attack strength $s$. If $G[y]$ is known, $s \, n_s$ is then formally given by

$$s \, n_s = \frac{1}{s!} G^{(s)}[0] \qquad (0.6)$$

where $G^{(s)}[y]$ is the $s$-th derivative of $G[y]$ with respect to $y$. $G^{(s)}[y]$ can be decomposed as

$$G[y] = n_1 \, y + \sum_{s=2}^{\infty} s' n_s \, y^s \equiv n_1 \, y + g[y] \qquad (0.7)$$

where the function $g[y]$ governs the attack-units' attack-strength distribution $n_s$ for $s \geq 2$. The next task is to obtain an equation for $g[y]$. This can be done in two ways. One could either write down the terms in $\left(g[y]\right)^2$ explicitly and then make use of Equation (0.3), or one could construct $g[y]$ by multiplying Equation (0.3) by $e^{-\omega s}$ and then summing over $s$. The resulting equation is:

$$\left(g[y]\right)^2 - \left(\frac{2-\nu}{1-\nu} N - 2n_1 \, y\right) g[y] + n_1^2 \, y^2 = 0 \qquad (0.8)$$

First we solve for $n_1$. From Equation (0.7), $g[1] = G[1] - n_1 = N - n_1$. Substituting $n_1 = N - g[1]$ into Equation (0.8) and setting $y = 1$, yields

$$g[1] = \frac{1-\nu}{2-\nu} N \qquad (0.9)$$

Hence 
$$n_1 = N - g[1] = \frac{1}{2-\nu} N \qquad (0.10)$$

To obtain $n_s$ with $s \geq 2$, we need to solve for $g[y]$. Substituting Equation (0.10) for $n_1$, Equation (0.8) becomes

$$\left(g[y]\right)^2 - \left(\frac{2-\nu}{1-\nu}N - \frac{2N}{2-\nu}y\right)g[y] + \frac{N^2}{(2-\nu)^2}y^2 = 0 \qquad (0.11)$$

Equation (0.11) is a quadratic equation for $g[y]$ which can be solved to obtain

$$\begin{aligned} g[y] &= \frac{(2-\nu)N}{4(1-\nu)}\left(1 - \sqrt{1 - \frac{4(1-\nu)}{(2-\nu)^2}y}\right)^2 \\ &= \frac{(2-\nu)N}{4(1-\nu)}\left(2 - \frac{4(1-\nu)}{(2-\nu)^2}y - 2\sqrt{1 - \frac{4(1-\nu)}{(2-\nu)^2}y}\right). \end{aligned} \qquad (0.12)$$

Using the expansion[vi]

$$(1-x)^{1/2} = 1 - \frac{1}{2}x - \sum_{k=2}^{\infty}\frac{(2k-3)!!}{(2k)!!}x^k, \qquad (0.13)$$

we have

$$g[y] = \frac{(2-\nu)N}{2(1-\nu)}\sum_{k=2}^{\infty}\frac{(2k-3)!!}{(2k)!!}\left(\frac{4(1-\nu)}{(2-\nu)^2}y\right)^k. \qquad (0.14)$$

Comparing the coefficients in Equation (0.14) with the definition of $g[y]$ in Equation (0.7), the probability of finding an attack unit of attack strength $s$ is given by:

$$\frac{s n_s}{N} = \frac{(2-\nu)}{2(1-\nu)}\frac{(2s-3)!!}{(2s)!!}\left(\frac{4(1-\nu)}{(2-\nu)^2}\right)^s. \qquad (0.15)$$

It hence follows that the average number of attack units of attack strength $s$ is

---

[vi] The 'double factorial' operator $!!$ denotes the product: $n!! = n(n-2)(n-4)\ldots$

$$n_s = \frac{(2-\nu)}{2(1-\nu)}\frac{(2s-3)!!}{s(2s)!!}\left(\frac{4(1-\nu)}{(2-\nu)^2}\right)^s N$$

$$= \frac{(1-\nu)^{s-1}(2s-2)!}{(2-\nu)^{2s-1}(s!)^2} N \tag{0.16}$$

The $s$-dependence of $n_s$ is implicit in Equation (0.16), with the dominant dependence arising from the factorials. Recall Stirling's series for $\ln[s!]$:

$$\ln[s!] = \frac{1}{2}\ln[2\pi] + \left(s+\frac{1}{2}\right)\ln[s] - s + \frac{1}{12s} - \cdots. \tag{0.17}$$

Retaining the few terms shown in Equation (0.17) is in fact a very good approximation, giving an error of $< 0.05\%$ for $s \geq 2$. This motivates us to take the logarithm of both sides of Equation (0.16) and then apply Stirling's formula to each log-factorial term, as in Equation (0.17). We follow these mathematical steps (which were derived in the M.Phil. thesis of Larry Yip, Chinese University of Hong King, who was supervised by Prof. Pak Ming Hui). We hence obtain

$$\ln(n_s) \approx \ln\left(\frac{(1-\nu)^{s-1}}{(2-\nu)^{2s-1}}N\right) + \left(2s-\frac{3}{2}\right)\ln(2s-2) + \ln\left(e^2\right) - \frac{1}{2}\ln(2\pi) - (2s+1)\ln(s)$$

$$\approx \ln\left(\frac{e^2 4^s (1-\nu)^{s-1}}{2^{\frac{3}{2}}\sqrt{2\pi}(2-\nu)^{2s-1}}N\right) + \left(2s-\frac{3}{2}\right)\ln(s) - \left(3s-\frac{3}{2}\right)\frac{1}{s} - (2s+1)\ln(s)$$

Combining the terms on the right-hand side into a single logarithm, it follows that

$$n_s \approx \left(\frac{(2-\nu)e^2}{2^{3/2}\sqrt{2\pi}(1-\nu)}\right)\left(\frac{4(1-\nu)}{(2-\nu)^2}\right)^s \cdot \frac{(s-1)^{2s-3/2}}{s^{2s+1}}N. \tag{0.18}$$

The $s$-dependence at large $s$ can then be deduced from Equation (0.18):

$$n_s \square N\left(\frac{4(1-\nu)}{(2-\nu)^2}\right)^s s^{-5/2}. \tag{0.19}$$

For small values of $\nu$, the dominant dependence on $s$ is therefore

$$n_s \sim s^{-5/2} \quad \text{hence} \quad n_s \sim s^{-2.5} \tag{0.20}$$

We have therefore shown analytically that the distribution of attack strengths will follow a power-law with a coefficient 2.5 (i.e. 5/2). As discussed earlier, we assume that any particular attack unit could be involved in an event in a given time interval, with a probability $p$ which is independent of its attack size. Hence these power-law results which we have derived for the distribution of attack strengths, will also apply to the distribution of attacks of severity $x$. (Recall that the attack strength $s$ is a measure of the number of casualties in a typical event, and that the severity $x$ of an event is measured as the number of casualties). In other words, the same power-law exponent 2.5 derived in Eq. (0.20), will *also* apply to the distribution of attacks having severity $x$.

**Hence our model predicts that any guerilla-like war which is characterized by an ongoing process of coalescence and fragmentation of attack units, and hence an ongoing re-distribution of the total attack strength, will have the following properties:**

(i) **The distribution of events with severity $x$ will follow a power-law. This finding is consistent with the behaviour observed for the aggregated data in the Iraq and Colombia wars (see Figure 1 of the paper).**

(ii) **The power-law distribution will, in the steady-state (i.e. long-time) limit, have a coefficient of 2.5. This is precisely the value to which the results for Colombia and Iraq currently seem to be heading (see Figure 2 of the paper).**

In the case of the Iraq war, we can go one step further by providing a simple generalization of the above model in order to offer an explanation for the evolution of the power-law coefficient throughout the war's entire history (recall Figure 2 of paper). The above model is characterized by the probability $\nu$ together with the mechanism for attack-unit coagulation and fragmentation. This value $\nu$ was chosen to be *independent* of the attack strength of the individual attack units involved. In this modification, we will keep the essential structure of the model, but we will add the modification that an attack unit will fragment with a probability which depends on its attack strength, and will coalesce with another attack unit with a probability depending on the attack strengths of the two attack units involved. With probability $\nu$ the randomly-chosen attack unit $i$ (chosen with probability proportional to the attack strength) will fragment into attack units of attack strength 1, with a probability $f[s_i]$ which depends on $s_i$. With probability $(1-\nu)$ the attack unit of attack strength $s_i$ coalesces with another randomly-chosen attack unit $j$ having attack strength $s_j$, with probability $f[s_i]f[s_j]$. They remain separated otherwise. With the choice $f[s]=1$ the original model is recovered. Analytically, this particular formulation of the fragmentation and coagulation process can be readily treated by the generating function approach discussed earlier, as will be demonstrated below.

Before proceeding, we discuss why this probabilistic 'attack-unit-formation' process may indeed mimic certain aspects of guerilla warfare. One such aspect is the effect of the arrival of opposing troops in the area. Imagine that at a given timestep and with a given probability $\nu$, the opposing army arrives in the vicinity of a given attack unit of attack strength $s_i$. If the overall conflict is such that the opposing army has the guerrillas/insurgents on the run, then this might suggest to the members of the insurgent attack unit that they should separate and move away from the area. However, if the state of the conflict is such that the guerilla/insurgent force feels powerful, they are unlikely to just disband and run if they have a significant attack strength. Instead they will possibly stand their ground and fight. Hence their probability of fragmentation is likely to be a decreasing function of their attack strength. By contrast, with probability $(1-\nu)$, no opposing troops arrive in the vicinity of the attack unit. With probability $f[s_i]$ ($f[s_j]$) the attack unit $i$ ($j$) decide to join forces. Thus, the two attack units will coalesce with probability $f[s_i]f[s_j]$. Again, this need for coalescing is likely to be less if the two attack units involved already feel powerful. Hence we would expect the probability of coalescence of the two attack units to be a decreasing function of their attack strengths. It is therefore quite plausible that -- depending on the state of the war from the insurgent force's perspective -- the probabilities of fragmentation and coalescence should depend on $f[s_i]$ ($f[s_j]$), i.e. they depend on the attack strengths of the attack units involved.

Analytically, the master equations for the specific example case in which $f[s] \sim s^{-\delta}$ can readily be written down:

$$\frac{\partial n_s}{\partial t} = -\frac{\nu s^{1-\delta} n_s}{N} + \frac{(1-\nu)}{N^2} \sum_{s'=1}^{s-1} (s')^{1-\delta} n_{s'}(s-s')^{1-\delta} n_{s-s'} - \frac{2(1-\nu)s^{1-\delta} n_s}{N^2} \sum_{s'=1}^{\infty} (s')^{1-\delta} n_{s'} \quad \text{for } s \geq 2 \quad (0.21)$$

$$\frac{\partial n_1}{\partial t} = \frac{\nu}{N} \sum_{s'=2}^{\infty} (s')^{2-\delta} n_{s'} - \frac{2(1-\nu)n_1}{N^2} \sum_{s'=1}^{\infty} (s')^{1-\delta} n_{s'} \quad (0.22)$$

with the physical meaning of each term being similar to that for Equations (0.1) and (0.2). The steady state equations become

$$s^{1-\delta} n_s = A \sum_{s'=1}^{s-1} (s')^{1-\delta} n_{s'}(s-s')^{1-\delta} n_{s-s'} \quad (0.23)$$

$$n_1 = B \sum_{s'=2}^{\infty} (s')^{2-\delta} n_{s'} \quad (0.24)$$

The constant coefficients $A$ and $B$ are given by

$$A = \frac{1-\nu}{N\nu + 2(1-\nu)\sum_{s'=1}^{\infty}(s')^{1-\delta}n_{s'}} \qquad \text{and} \qquad B = \frac{N\nu}{2(1-\nu)\sum_{s'=1}^{\infty}(s')^{1-\delta}n_{s'}}$$

Setting $\delta = 0$ in Equations (0.23) and (0.24) recovers Equations (0.3) and (0.4) for the original model. A generating function

$$G[y] = \sum_{s'=0}^{\infty}(s')^{1-\delta}n_{s'}\, y^{s'} = n_1\, y + g[y] \qquad (0.25)$$

can be introduced where $g[y] = \sum_{s'=2}^{\infty}(s')^{1-\delta}n_{s'}\, y^{s'}$ and $y = e^{-\omega}$. The function $g[y]$ satisfies a quadratic equation of the form

$$\big(g[y]\big)^2 - \left(\frac{1}{A} - 2n_1\, y\right)g[y] + n_1^2\, y^2 = 0 \qquad (0.26)$$

which is a generalization of Equation (0.8). Using $n_1 + g[1] = \sum_{s'=1}^{\infty}(s')^{1-\delta}n_{s'}$ and Equation (0.26), $n_1$ can be obtained as

$$n_1 = \frac{(1-\nu)^2 - \nu^2 A^2 N^2}{4(1-\nu)^2 A} \qquad (0.27)$$

Solving Equation (0.26) for $g[y]$ gives

$$g[y] = \frac{1}{4A}\left(1 - \sqrt{1 - 4n_1 A\, y}\right)^2 \qquad (0.28)$$

Following the steps leading to Equation (0.19), we obtain $n_s$ in the modified model:

$$n_s \square N \left( \frac{4(1-\nu)\left((1-\nu) + \dfrac{N\nu}{\sum_{s'=1}^{\infty}(s')^{1-\delta}n_{s'}}\right)}{\left(\dfrac{N\nu}{\sum_{s'=1}^{\infty}(s')^{1-\delta}n_{s'}} + 2(1-\nu)\right)^2} \right)^{s} s^{-(5/2-\delta)} \qquad (0.29)$$

For $\delta = 0$, $\sum_{s'=1}^{\infty}(s')^{1-\delta}n_{s'} = N$ and hence Equation (0.29) reduces to the result in Equation (0.19) for the original model. For $\delta \neq 0$, it is difficult to solve explicitly for $n_s$. However the summation simply gives a constant, and thus for small $\nu$ the dominant dependence on the attack strength $s$ is $n_s \sim s^{-(5/2-\delta)}$ and hence equivalently $n_s \sim s^{-(2.5-\delta)}$.

**Most importantly, we can see that by decreasing $\delta$ from $0.7 \rightarrow 0$ (i.e. by increasing the relative fragmentation/coalescence rates of larger attack units) we span the entire spectrum of power-law exponents observed in the Iraq war from the initial value of 1.8, up to the current tendency towards 2.5. This effect of decreasing $\delta$ from $0.7 \rightarrow 0$ corresponds in our model to a relative increase in the tendency for larger attack units to either fragment or coalesce at each timestep. In other words, decreasing $\delta$ mimics the effect of decreasing the relative robustness or 'lifetime' of larger attack units.**

Going further, we note that these theoretical results are consistent with, *and to some extent explain*, the various power-law exponents found for:

(1) Conventional wars. The corresponding power-law exponent 1.8 can now be interpreted through our generalized model with $\delta \approx 0.7$, as a tendency toward building larger, robust attack units with a fixed attack strength as in a conventional army -- as opposed to attack units with rapidly fluctuating attack strengths as a result of frequent fragmentation and coalescence processes. There is also a tendency to form a distribution of attack units with a wide spectrum of attack strengths – this is again consistent with the composition of 'conventional' armies from the past.

(2) Terrorism in G7 countries. The corresponding power-law exponent 1.7 can be interpreted through our generalized model with $\delta \approx 0.8$, as an even stronger tendency for robust units (e.g. terrorist cells) to form. There is also an increased tendency to form larger units – or rather, to operate as part of a large organization.

(3) Terrorism in non-G7 countries. The corresponding power-law exponent 2.5 can be interpreted through our model with $\delta = 0$, as a tendency toward more transient attack units (e.g. terrorist cells) whose attack strengths are continually evolving dynamically as a result of an on-going fragmentation and coalescence process. Unlike a conventional army, there will be a tendency to form smaller attack units rather than larger ones.

Interestingly, we can now discuss the evolution of the wars in Colombia and Iraq in these terms:

*War in Colombia.* At the beginning of the 1990's, the power-law exponent was very high (3.5). Then over the following 15 years, it gradually lowered to the present value and appears to be tending toward 2.5. Using our model, the interpretation is that the war at the beginning of the 1990's was such that the guerrillas favoured having small attack units. This is possibly because they lacked communications infrastructure, and/or did not feel any safety in larger numbers. The decrease toward the value 2.5, suggests that this has changed – probably because of increased infrastructure and communications, enabling attack units with a wide range of attack strengths to build up.

*War in Iraq.* At the beginning of the war in 2002, the power-law exponent was quite low (1.8) and was essentially the same value as conventional wars. This is consistent with the war being fought by a conventional Iraqi army against the Coalition forces. There is then a break in this value after a few months (i.e. the war ended) and following this, the power-law exponent gradually rises towards 2.5. This suggests that the insurgents have been increasingly favouring more temporary attack units, with an increasingly rapid fragmentation-coalescence process. This finding could be interpreted as being a result of increased success by the Coalition Forces in terms of forcing the insurgents to fragment. On the other hand, it also means that the Iraq War has now moved to a value, and hence character, which is consistent with generic non-G7 terrorism.

**PART 2: Supplementary tables and figures which help confirm the robustness of our results**

| Estimates of power-law coefficients for the entire time-series | | | | | | |
|---|---|---|---|---|---|---|
| | $\alpha_1$ | Confidence Interval | Percentage of points inside interval | $\alpha_2$ | Confidence Interval | Adjusted $R^2$ |
| K | 3.1013 | +/- 0.02 | 0.95% | 2.8761 | +/- 0.0151 | 0.9835 |
| I | 3.04 | +/- 0.02 | 0.95% | 2.9717 | +/- 0.0211 | 0.9818 |
| KI | 2.93 | +/- 0.015 | 0.95% | 3.0061 | +/- 0.0179 | 0.9796 |
| $CK_{min}$ | 2.07 | +/- 0.005 | 0.90% | 2.1279 | +/- 0.0057 | 0.9336 |
| $CK_{max}$ | 2.02 | +/- 0.003 | 0.90% | 2.0966 | +/- 0.0043 | 0.9496 |

**Supplementary Table 1** Shows two complementary estimates of the power-law coefficients for the variables K (reported deaths for Colombia), I (reported injuries for Colombia), KI (reported deaths plus injuries for Colombia), $CK_{min}$ (minimum reported civilian deaths for Iraq) and $CK_{max}$ (maximum reported civilian deaths for Iraq). Our first estimate ($\alpha_1$) uses $\alpha = 1 + n\left[\sum_{i=1}^{n}\ln\left(x_i/x_{min}\right)\right]^{-1}$ while our second estimate ($\alpha_2$) uses ordinary least-squares linear regression. The two results are always very similar and the results vary little as we vary the victimization measure for Colombia. For further discussion, see PART 3 of this Supplementary Information document.

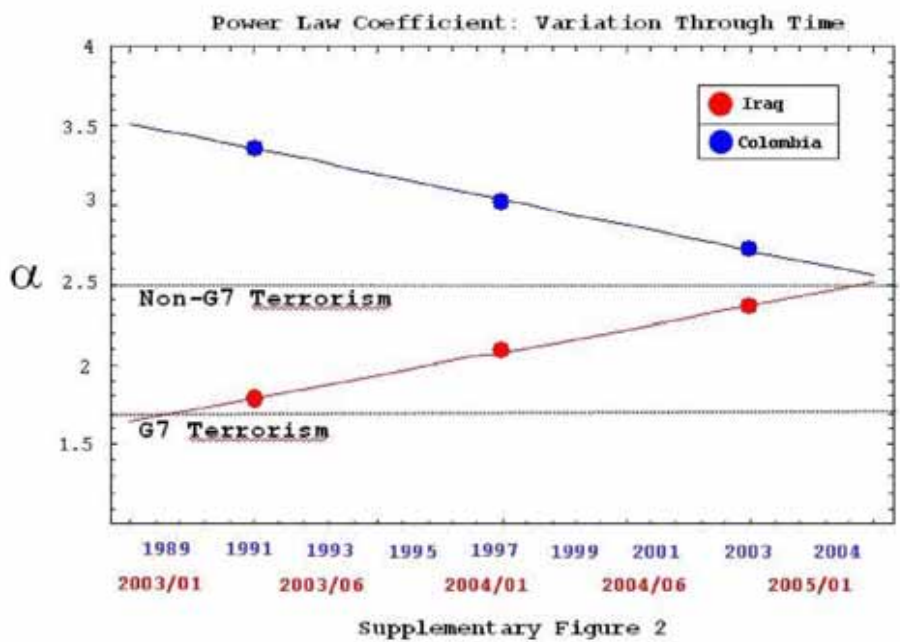

Supplementary Figure 2

**Supplementary Figure 2** The variation through time of the power law coefficients for three 2,500 day intervals displaced by 1,855 days for the Colombian data and three 365 day intervals displaced every 258 days for the Iraq data. Despite this change in size of the windows and how they slide across time both curves seem to be tending toward 2.5, as in Figure 2 of the paper. For further discussion, see PART 3 of this Supplementary Information document.

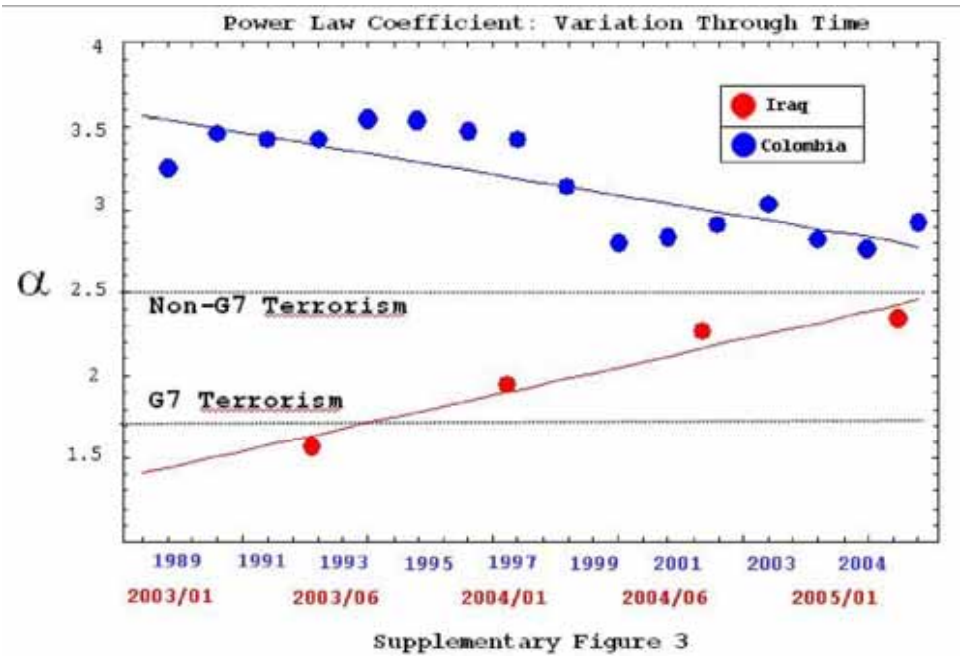

Supplementary Figure 3

**Supplementary Figure 3** The variation through time of the power law coefficients for two year intervals displaced every year for Colombia and 200 day intervals displaced every 200 days for Iraq. Again, they both seem to be tending toward 2.5, as in Figure 2 of the paper. For further discussion, see PART 3 of this Supplementary Information document.

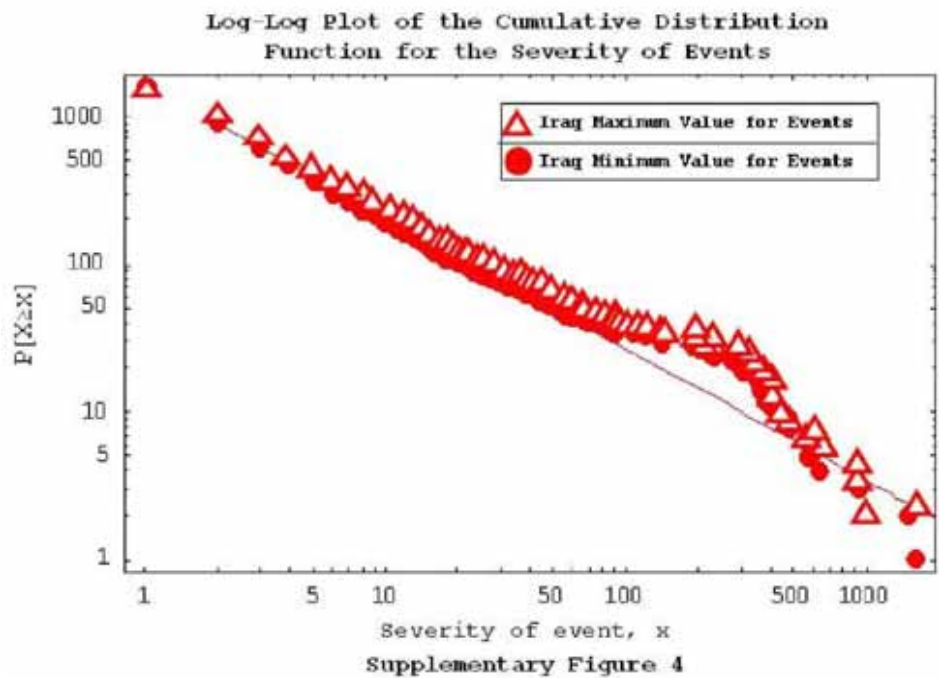

Supplementary Figure 4

**Supplementary Figure 4**   Log-Log plots of cumulative distributions $P(X \geq x)$ describing events greater than $x$, for the minimum possible value and maximum possible value of each event in the Iraq dataset.  The results are very much the same across the two measures. For further discussion, see PART 3 of this Supplementary Information document.

**PART 3:  Supplementary Notes on Methods**

If $p(x) = Cx^{-\alpha}$, then, $C = (\alpha-1)x_{\min}^{(\alpha-1)}$ and $\alpha = 1 + n\left[\sum_{i=1}^{n}\ln\left(\frac{x_i}{x_{\min}}\right)\right]^{-1}$.

We used two estimates for $\alpha$ (see Supplementary Information Table 1 in PART 2). We estimated $\alpha_1$ using $\alpha$ with $x_{\min}$ equal to the minimum value that satisfies the Kolmogorov statistic for the whole data set.  We estimated $\alpha_2$ using ordinary least-square regression analysis for all the values greater than $x_{\min}$.

We established $x_{\min}$ as the minimum value of $x$ where we could not reject the hypothesis that the data beyond $x_{\min}$ followed a power law using a Kolmogorov-Smirnov goodness of fit test at 95% confidence.

In order to estimate the error of $\alpha$, we used a minus-one jackknife resampling method.  We created the datasets obtained by removing one value from the original dataset. For each of these surrogates, we estimated $\alpha$ and $C$.  We then measured the maximum and minimum deviation from the mean of the parameters $\alpha$ obtained from the diverse jackknife datasets. We established an interval equal to the maximum distance from the mean of the sample times 5.  Since this is not an analytical solution, we present the percentage of values that fall into that interval.

**The Evolution of** $\alpha$:  To test the robustness of our findings in Figure 2 of the paper, we have repeated the calculation of $\alpha$ for several different sizes of the time windows. We also tried varying the way in which these windows slide forward in time.  All these changes barely affected our results.

As an example, in Part 2 of this document we have plotted the evolution of KI (deaths and injuries) for different time windows. For the Colombian data set we used a time window of 2,500 days moved every 1,855 days (see Supplementary Figure 2); then a moving time window of two years displaced every year (see Supplementary Figure 3), and finally a moving time window of two years, displaced every 50 days (Figure 1). For the Iraq data set we used a 365 day time window displaced every 258 days (Supplementary Figure 2), a 250 day interval displaced every 150 days (Supplementary Figure 3), and a 365 day time window displaced every 30 days (Figure 1). As can be seen, our results are essentially unchanged by these variations.

**Further general comment on robustness testing:** As a further test of the robustness of the results obtained in our paper, and in particular our main findings in Figures 1 and 2 of the paper, we ran the following variations of our calculations. For Colombia we used just killings and just injuries, rather than killings plus injuries as presented in the paper. For Iraq we used the maximum number of deaths rather than the minimum number of deaths as reported in the paper. These results are shown in Supplementary Figure 4. As can be seen, these variations do not affect our findings. This is reassuring, and is actually not too surprising since the power-law coefficient $\alpha$ provides a statistical measure of the structure of the events' time-series, rather than the absolute number of killings and/or injuries. Hence the power-law coefficient $\alpha$ will be unaffected by any constant scale factors which are introduced as a result of a fixed ratio of injuries to killings.

An additional, but perhaps even more important, advantage of focusing on $\alpha$, concerns possible over- or under-reporting of war casualties. In particular, $\alpha$ is insensitive to systematic over-reporting or under-reporting of casualties. This is because any systematic multiplication of the raw numbers by some constant factor, has no affect on the $\alpha$ value which emerges from the log-log plot.